# Improved accuracy and reproducibility of coronary artery calcification features using deconvolution

Yingnan Song[a], Ammar Hoori[a], Hao Wu[a], Mani Vembar[b], Sadeer Al-Kindi[c], Leslie Ciancibello[d], James G. Terry[e], David R. Jacobs, Jr[f], John Jeffrey Carr[e], David L. Wilson[a,g]*

[a] Department of Biomedical Engineering, Case Western Reserve University, Cleveland, OH, USA
[b] Philips Healthcare, Cleveland, OH, USA
[c] Department of Cardiology, University Hospitals, Cleveland, OH, USA
[d] Department of Radiology, University Hospitals, Cleveland, OH, USA
[e] Department of Radiology, Vanderbilt University Medical Center, Nashville, TN, USA
[f] University of Minnesota, Minneapolis, MN, USA
[g] Department of Radiology, Case Western Reserve University, Cleveland, OH, USA

**Abstract**

**Purpose:** Our long-range goal is to improve current whole-heart CT calcium score by extracting quantitative features from individual calcifications. We performed deconvolution to improve accuracy/reproducibility of small calcifications which challenge conventional CT calcium score scanning resolution.

**Approach:** We analyzed features of individual calcifications on repeated standard (2.5-mm) and thin (1.25-mm) slice scans from QRM-Cardio phantom, cadaver hearts, and CARDIA study participants. Preprocessing to improve resolution involved of Lucy-Richardson deconvolution with a measured PSF or 3D blind deconvolution where the PSF was iteratively optimized on high detail structures like calcifications in the images.

**Results:** Using QRM with inserts having known mg-calcium, we determined that both blind and conventional deconvolution improved mass measurements nearly equally well on standard images. Further, deconvolved thin images gave excellent recovery of actual mass scores, suggesting that such processing could be our gold standard. For CARDIA images, blind deconvolution greatly improved results on standard slices. Accuracy across 33 calcifications (without, with deconvolution) was (23%,9%), (18%,1%), and (-19%,-1%), for Agatston, volume, and mass scores, respectively. Reproducibility was (0.13,0.10), (0.12,0.08), and (0.11,0.06), respectively. Mass scores were more reproducible than Agatston scores or volume scores. Cadaver volumes showed similar improvements in accuracy/reproducibility and slightly better results with a measured PSF. For many other calcification features in CARDIA data, blind deconvolution improved reproducibility in 21 out of 24 features.

**Conclusions:** Deconvolution improves accuracy and reproducibility of multiple features extracted from individual calcifications in CT calcium score exam. Blind deconvolution is useful for improving feature assessments of coronary calcification in archived datasets.

**Keywords**: CT calcium score, cardiovascular disease, deconvolution, image processing

*David Wilson, E-mail: david.wilson@case.edu

## 1 Introduction

Cardiovascular disease is the most common cause of death in the United States, and coronary artery disease (CAD) is the most common type of heart disease[1,2]. CT coronary artery calcium (CAC) gives direct evidence of atherosclerotic coronary artery disease, which can be obtained via a fast, reliable, non-invasive and non-contrast-enhanced examining method. As coronary calcium is easy to reliably detect, it results in an examination with extremely high sensitivity/specificity. The exam is low cost creating an opportunity for screening of individuals in high-risk categories. At our institution (University Hospitals of Cleveland), the CT calcium score exam is currently free, with ~13,000 exams conducted annually.

A great number of studies have shown that CT calcium whole-heart Agatston score predicts risk of adverse cardiovascular events [3–9]. Agatston score, as an accurate marker of subclinical coronary artery disease, is more predictive than other single biomarkers, including lipids[5,10]. There is emerging evidence that shows that regional calcification is important[11,12]. Whole heart Agatston score is a traditional measure of calcium on a coronary CT calcium scan. However individual calcifications are important, since whole-heart Agatston can easily be dominated by the largest, densest, stable calcifications. From pathobiology and clinical observations, it is likely that small, spotty, low-density calcifications will provide better evidence of disease progression than whole-heart Agatston score. In fact, a recent study suggested that patients with very high calcification densities (>1000 Hounsfield Unit (HU)) have reduced risk[13], contrary to Agatston, suggesting room for improvement [11].

Other common assessments in CT calcium score exams are volume and mass scores [14–16]. While the volume score is the total number of calcified arterial voxels, the mass score is the accumulation of an actual mineral mass expressed in milligrams [17], which has inherent advantages in the presence of partial volume effects. Various studies showed that mass score is less variable compared to Agatston score and volume score [18–21]. Typical Agatston, volume, and mass scores are given as single numbers for the entire heart. In our work, we are particularly interested in identifying features of individual calcifications (e.g. number of lesions detected in 3D, maximum HU value, maximum mass score, and more), including small calcifications, suggesting a need for corrections aimed at obtaining more accurate and reproducible measurements [22]. Several papers have identified that feature reproducibility is an important requirement for machine learning [23–27].

There is related image processing work. CT images are blurred due to the focal spot size, reconstruction filter, motion of the gantry during sampling of a projection, etc. To address blurring, Liang et al. applied deconvolution image restoration to reduce the blooming in cardiac CT images [28], while Hijarrubia et al. claimed that deconvolution restores small high-density structures in micro-CT, which enhanced the visualization of calcification [29]. In a preliminary report, Richards et al developed a motion point spread function (PSF) based method on stenosis estimation [30,31] and Wang et al. proposed using different weight values for smooth region and edge region during the deblurring process to suppress ringing[32]. Apart from deconvolution, partial volume corrections can be applied, wherein the assumption is that voxels at an interface are "averaging" a mixture of a calcification and soft tissue. Recently, we have developed a method for partial volume correction [33], and Šprem et al. reported a similar approach [34].

In this study, we analyzed and compared the use of appropriate deconvolution correction methods on CT calcium score exam images obtained from phantoms with known calcification mass, heavily calcified cadaver hearts, and the Coronary Artery Risk Development in Young Adults (CARDIA) cohort study. In experiments, we analyzed accuracy and reproducibility of features of individual calcifications with and without deconvolution of standard slice thickness (2.5-mm) scans. As the 3D PSF of a CT imaging system is not always known, we compared conventional deconvolution with known PSF and blind deconvolution which estimates the PSF from the image itself. In addition to Agatston, mass, and volume of individual calcifications, we analyzed reproducibility of various features for potential use in machine learning analyses of the major adverse cardiovascular event risk.

## 2     Methods

*2.1 CT imaging*

To estimate the reproducibility of calcification features, CT imaging of phantoms, cadavers was performed as follows: The QRM-Cardio phantom with known mass values and 10 cadaver hearts were scanned at three different angles (−15, 0, +15 deg) using Philips IQon spectral CT (Philips Healthcare, Best, The Netherlands) in standard CT mode. The scan protocol used was standard slice scan (SS, slice thickness: 2.5-mm contiguous, exposure: 55 mAs, and filter type: CB) and thin slice scan (TS, slice thickness: 0.67-mm contiguous, exposure: 200 mAs, and filter type: CB). The reconstruction method was "cone beam" with 0.49-mm in-plane voxel size.

We also analyzed 20 participants scan/rescan data from CARDIA Y20. CARDIA dataset recruited young black and white men and women at ages 18-30 years in 1985 and follow-up at 5 year interval [12]. At Y20, repeated scans were acquired at about a 5-minute interval. Images were acquired at standard slice scan (SS, slice thickness: 2.5 contiguous, exposure: 100 mAs, filter type: body filter) and thin slice (TS, slice thickness: 1.25-mm contiguous, exposure: 100 mAs, filter type: body filter) using a similar pixel size (0.68-mm in-plane). Using high quality interpolation (3D interpolation based on cubic convolution), we converted slice thicknesses of all TS scans of cadaver heart and phantom to be consistent with the CARDIA dataset acquisition (slice thickness: 1.25-mm).

*2.2 Calcification features*

As described in the Introduction, we wanted to analyze individual calcifications. We computed traditional features (Agatston, volume, and mass) using methods described in the literature. Agatston score was calculated according to the maximum HU value of the calcified region ($\geq$ 3 connected voxels over 130 HU) in axial 2D CT images. The total Agatston score was calculated by multiplying a weighting factor with the 2D area of individual calcified region in each axial slice, where the weighting factor was determined by the maximum HU value [35]. Volume score was calculated as the number of connected voxels meeting detection criterion. Mass score was calculated using the calibration curve generated from mean HU value of two calibration QRM phantom inserts, which represented the relationship between HU value of the voxel and calcium concentration. We manually segmented the calcification masks accepting connected voxels over 130 HU and saved the individual calcified region as a binary mask. We determined calibrations for CT scanner using two calibration calcium inserts in QRM phantom (shown in Figure 1). For mass score assessment, we assumed that the HU value varies linearly with density of hydroxyapatite (HA) with an offset in water inserts and used water equivalent material inserts and 200mgHA/cm$^3$ calibration inserts to generate the curve. Then we selected out the volume at the center of the calibration inserts and calculated the mean HU value of the voxels to generate calibration factor k. Mass score was evaluated using equation (1). To validate our results, we also analyzed calcium mass scores using a commercial software for comparison.

$$\text{Mass score} = \sum \text{k} * HU_{voxel} \qquad (1)$$

where k is the calibration factor generated by the two calibration inserts in the phantom, and *HUvoxel* represents the HU value of the voxel selected. In our experiment, calibration factor k in the phantom images was 0.71 (mgHA/cm$^3$)/HU for 120 kVp images.

We also evaluated Individual Calcification MORphological and mass featurEs (ICmore). ICmore is a comprehensive analysis of the individual calcification in CT calcium score exam, which includes HU value related features like mean/median HU value and morphological features like volume and first moment. We divided these into two groups: Accumulated Heart and Artery (AHA) and 3D calcification features. AHA features are the analysis of the whole heart, like whole-heart Agatston, mass and volume score, while 3D features look into individual calcifications, such as mean, max and standard deviation of mass score. Details can be found in Figure 10.

*2.3 PSF measurement and deconvolution*

The generation of CT images can be modeled as a linear space-invariant system as in Equation 2.
$$I(x,y,z) = h(x,y,z) * I_{origin}(x,y,z) + n(x,y,z) \qquad (2)$$
where $I(x,y,z)$ is the output image, $h(x,y,z)$ is the point spread function (PSF) of the system, $I_{origin}(x,y,z)$ represents the real image structure, which is the input of the system, and $n(x,y,z)$ is additive noise.

This operation in equation (2) blurred the image details, especially the edges and small objects, thus reducing the accuracy of calcium mass score calculation. To recover the real image, we need to estimate the additive noise and the PSF. We measured the additive noise by selecting an area composed of the same material expected to be homogeneous on the phantom. To measure the 3D PSF, we used tiny plastic sphere beads (diameter: 0.2 mm) which are smaller than one voxel and were embedded in OCT gel as phantom to provide adequate contrast. We acquired the PSF from 9 beads and averaged them. Both SS (slice thickness: 2.5-mm) and TS (slice thickness: 1.25-mm) scans were obtained, and to match slice thicknesses, we interpolated the SS volume to TS using 3D interpolation based on cubic convolution. To measure PSFs, we manually selected beads with a bounding box, subtracted the background, and fit a 10-parameter 3D Gaussian model (3 σ's; scaling parameter; x, y, and z offsets and peak intensities) by minimizing the squared error.

The Lucy-Richardson method is an iterative procedure for recovering an underlying image that has been blurred by a known PSF [30,31]. This algorithm uses both the PSF and noise characteristic as priors in the processing, enabling high contrast images with reduced noise as compared to other deconvolution methods. We used the damped Richardson-Lucy deconvolution algorithm implemented in MATLAB (MathWorks). The 3D PSF consisted of a $6.24 \times 6.24 \times 8.71$ mm$^3$ noise-free 3D array generated from the Gaussian PSF model. In processing, we set some adjustable parameters (values) to be as follows: the maximum number of iterations (80 times), the threshold for damping (0 HU), and subsampling (exact size). In addition, we provided to the algorithm an "additive noise array" obtained from calcification-free regions. Similar to Wang et al. [32] who used weight values for deblurring, we created a weighted array consisted of ones in a relatively large region around each calcification and zeros elsewhere. Anisotropic diffusion filtering was applied after deconvolution to depress noise. By running the diffusion filter with a 3D edge-seeking diffusion coefficient for a certain number of iterations (5 times), the image can be evolved towards a piecewise constant image with the boundaries between the constant components being detected as edges. We applied 3D Lucy-Richardson deconvolution corrections using a measured point spread function on phantom and cadaver heart volume of SS images (Standard Slice thickness after Deconvolution, SSD) and TS images (Thin Slice thickness after Deconvolution, TSD). We used a 3D blind damped Lucy-Richardson deconvolution algorithm on CARDIA participants' volume on SS image (Standard Slice thickness after Blind Deconvolution, SSBD). It utilizes the maximum likelihood algorithm and the initial estimate of the PSF. The initial estimate for the PSF was set to the same value as that which we measured above.

*2.4 Evaluation method*

We analyzed QRM phantom with 9 calcification inserts of three sizes (1 mm, 3 mm, 5 mm in diameter) and three densities (200 mgHA/cm$^3$, 400 mgHA/cm$^3$, 800 mgHA/cm$^3$) and two calibration inserts (water and 200 mgHA/cm$^3$). Densities and sizes of inserts were given in datasheet. We also analyzed 92 individual calcifications in 10 cadaver hearts and 33 individual calcifications in 20 CARDIA participants. For CT calcium score, accuracy was defined as the average percent signed difference {mean [(measurement - gold standard)/gold standard]} and reproducibility was defined as the coefficient of variation in the angled or repeated scans. We could easily get gold standard in QRM phantom with given values and created ground truth data in cadaver hearts study and CARDIA participants, details were written below in results. In addition, we analyzed reproducibility of ICmore across repeated scans in CARDIA participants using intraclass correlation coefficient (ICC).

## 3   Results

Significant reduction in peak intensity was observed when imaging calcification inserts in the QRM phantom (Figure 1). The effect was profound on the smallest calcification insert (1-mm diameter, 1-mm height). For the smallest 800 mgHA/cm$^3$ insert, the peak HU value (152 HU) was reduced by 90% as compared to the actual HU value based on hydroxyapatite concentration (1268 HU). The HU value barely exceeded the standard threshold for calcification detection (i.e., 130 HU).

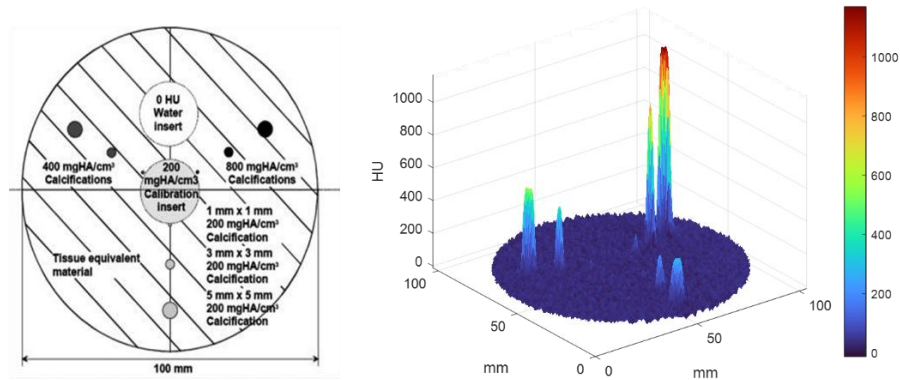

**Figure 1. QRM phantom datasheet.** Left panel: QRM phantom with nine calcification inserts of three sizes (1 mm, 3 mm, 5 mm in diameter) and three densities (200 mgHA/cm$^3$, 400 mgHA/cm$^3$, 800 mgHA/cm$^3$) and two calibration inserts (water and 200 mgHA/cm$^3$). More information can be found at website: https://www.qrm.de/en/products/cardiac-calcification-phantom/. Densities and sizes of inserts were given in the datasheet so that we can determine gold standard measurements. The density, volume and mass scores were given, so we can infer the theoretical HU value according to the calibration curve. Theoretical HU values were 302 HU, 623 HU and 1268 HU, respectively. Right panel: Surface plot of a CT image slice (standard slice thickness) containing the calcification inserts. Even though the calcification density was fixed along a radius, the smaller inserts showed significantly less peak signal.

PSF measurements, we found insignificant changes in parameters in different locations and averaged parameters across nine beads to get an average PSF. From 3D PSF measurements obtained in SS and TS scans, we got $\sigma_x = \sigma_y = 0.58 \pm 0.08$ mm in the transverse plane, and $\sigma_z$'s are $1.5 \pm 0.5$ mm(slice thickness: 2.5-mm) and $0.84 \pm 0.34$ mm(slice thickness: 1.25-mm), respectively. There was very good agreement between bead measurements obtained with a clinical scan in the transverse plane to those reported previously using Catphan 600 phantom [36].

We obtained images of the QRM phantom and visually evaluated the effect of deconvolution (Figure 2). We took the TS as the reference in this set of images, we could see significant brightness reduction in SS especially in the smallest calcium inserts with high density (red circle). However, after correction, SSBD and SSD recovered image contrast significantly. In the corrected clinical scans, the center HU value increased by 68% (164

HU to 276 HU) and 74% (164 HU to 286 HU) after SSBD and SSD, respectively, and the peak HU difference (286 HU and 291 HU) between TS and SSD was only 2% in this insert.

Then we quantitatively compared calcification mass scores of the QRM phantom obtained from SS and TS scans, with and without deconvolution (Figure 3, Table S1). The actual "ground truth" phantom mass scores (dashed lines) were obtained from the concentration of hydroxyapatite (HA) and the volume of the inserts. Figure 3(a) describes the clinical relevant density inserts in different sizes while Figure 3(b) describes the subclinical inserts with different densities. SSD correction improved accuracy and reproducibility decreased from 0.12 to 0.04 across the angles for all cases. Also processing improved accuracy in all phantom inserts as accuracy improved from 13% to 8%. All small inserts in the QRM phantom were detected by our standard criterion and provided reasonable results compared to ground truth. TSD yielded excellent accuracy results across all the measurements with values within about 2.1% of the ground truth. Reproducibility showed similar results as the coefficient of variation reduced from 1.8% to 1.4% after correction. Obtaining the true mass score of each individual calcification within the hearts is a laborious task. As a result, in the cadaver heart experiments, we assumed TSD to be the "gold standard" mass score, even if it was not the real physical value, as it was the most accurate practical measurement.

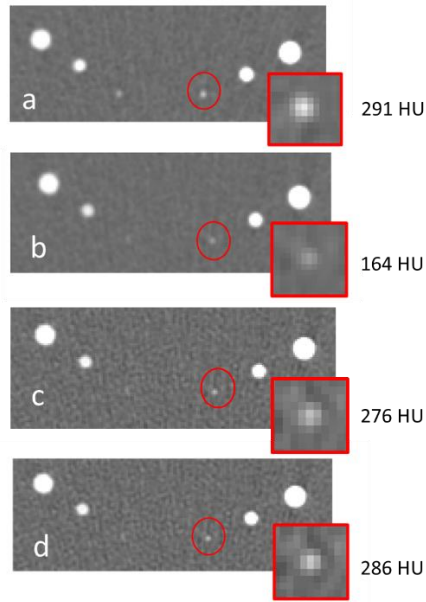

**Figure 2. Improvements in image quality with deconvolution (phantom).** We compared the 0 degree images at a: Thin Slice thickness (TS), b: Standard Slice thickness (SS), c: Standard Slice thickness after Blind Deconvolution (SSBD) and d: Standard Slice thickness after Deconvolution (SSD). In each image, the left set was the 400mgHA/cm$^3$ and the right set was 800mgHA/cm$^3$ with 3 different sizes. Here we took the smallest insert (1mm in diameter and height) in 800mgHA/cm$^3$ set as circled example to illustrate the deconvolution influence. The sub-images with red border were the zoomed in calcifications in circled locations and the center HU value of this insert is labeled on right. The edge of the calcium was significantly enhanced, making the tiny cylinder calcification in phantom detectable under the 130 HU threshold. Also, 3D deconvolution (d) that enhanced the center HU value as the calcified region was much more similar to the TS one (a).

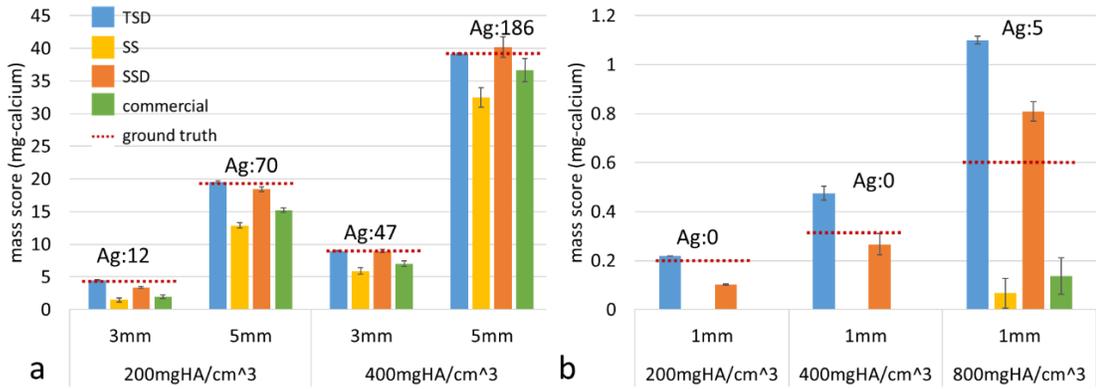

**Figure 3. Comparisons to actual QRM-Cardio phantom values.** Evaluations (Thin Slice thickness after Deconvolution(TSD), Standard Slice thickness(SS), Standard Slice thickness after Deconvolution(SSD) and commercial software) were compared to actual mass scores (dashed line). Error bars were from scan with angles. Ag stands for Agatston score. (a) showed results in clinical relevant calcification inserts and (b) showed results in smallest calcification inserts. In (a), TSD excellently agreed with actual, leading us to use this as a gold standard in heart imaging. Deconvolution significantly improved results from standard images, especially in (b), we can see that small calcification inserts can be detected and giving reasonable results after deconvolution.

We scanned and analyzed calcifications in 10 cadaver hearts and most of them were isolated spotty calcifications with a mass score of less than 20 mg-calcium. As for the cadaver images (Figure 4), we could visually see that SSD corrected the calcifications to be more similar to TS scan, and contrast was restored as the edge of zoomed calcification get much clearer. In Figure 5 and Table S2, we selected four calcifications from a cadaver heart to make detailed comparisons of measurements, with and without correction. In SS, scores were underestimated as compared to gold standard, but after correction, SSD results were the best. Variation in angled measurements was reduced with SSD in some instances. In Figure 6, we presented a modified Bland-Altman plot of all 92 calcification mass score evaluations. In all cases, we compared measurements to the gold standard measurement (TSD). The plot shows bias in a measurement when the average horizontal curve is different from the axis at zero; shows the spread of measurements, indicating precision; and it shows reproducibility as each datum includes the mean and standard deviation across measurements at different angles. We also include SSBD results in this plot. Considering bias and precision results, methods for processing can be ordered as SSD > SSBD > commercial software ≈ SS using our analysis software. For example, biases were -2.3, -4.32, -9.42 and -10.59 mg-calcium, respectively. Similarly, standard deviations were 3.96, 4.53, 8.35 and 9.07 mg-calcium, respectively.

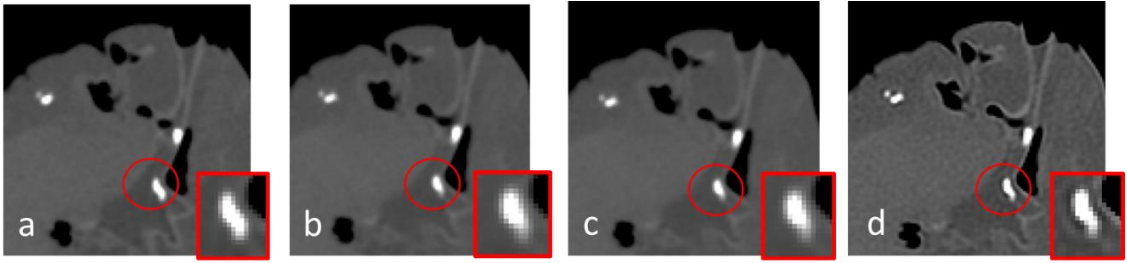

**Figure 4. Improvements in image quality with deconvolution (cadaver heart).** We showed 0 degree images at TS (a), SS (b), SSBD (c) and SSD (d) deconvolution correction. The sub-images with red border were the zoomed in calcifications in circled locations. The mass scores were 15.43 mg-calcium, 11.59 mg-calcium,13.23 mg-calcium and 14.52 mg-calcium respectively. Deconvolution corrected images and had less blurring artifact, the improved image(d) was similar to the TS (a).

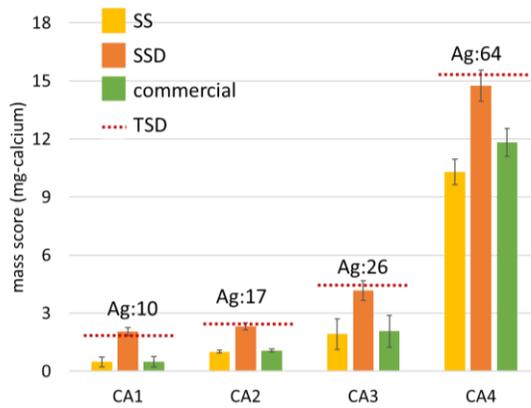

**Figure 5. Comparisons to thin slice deconvolution cadaver values.** Image conditions (SS, SSD and commercial software) are compared to TSD (dashed). Deconvolution significantly improved results from standard images.

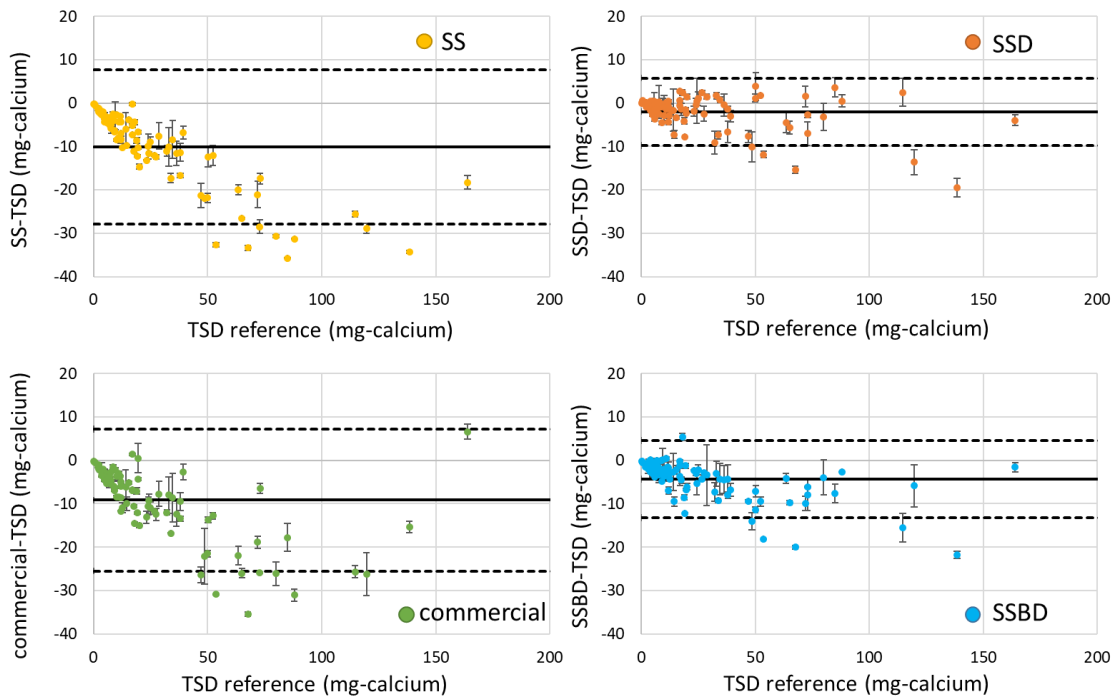

**Figure 6. Modified Bland-Altman plot of the calcium mass score in 10 cadaver hearts.** The X-axis is reference mass score, which is TSD results. The Y-axis is the difference between the measurement and reference. The solid black line describes bias all over the measurement in the figure while the dashed line describes the limits of agreement. Ideal measurements come with bias close to 0 and narrow range of limits of agreement. SSD performs better than the commercial software, improves the accuracy by 78% and precision by 52% comparing to SS, and SSBD shows similar performance comparing to the SSD. Large calcifications are less reproducible and underestimated a lot even in the corrected analysis.

In CARDIA images, we found that SSBD improved accuracy (Figure 7, Table S3, Figure 8, Figure 9, Figure 10) comparing to our reference Thin Slice thickness after Blind Deconvolution (TSBD). For Agatston, volume, and mass score, the calculated average percent signed difference were (23%, 9%), (18%, 1%), and (-19%, -1%), respectively for SS and SSBD. Assuming TSBD to be the reference, difference between measurement and reference suggested standard slice with deconvolution processing is accurate comparing to reference (one-sample t-test, p=0.47, p=0.56 and p=0.63 for Agatston score, volume score and mass score, respectively). For reproducibility, we calculated average coefficient of variation, results were (0.13, 0.1), (0.12, 0.08), and (0.11, 0.06), for SS and SSBD, for Agatston, volume, and mass score, respectively, showing the most reproducible assessment was mass score following deconvolution. Similar results were found in cadaver hearts as well as mass score improved accuracy from 26% to 3% and reproducibility from 0.14 to 0.08. We also investigated in the reproducibility of calcification features in CARDIA scan-rescan. We observed improvement of calcification number reproducibility using blind deconvolution (Figure 9). Deconvolution reduced differences in scan-rescan from 9 to 2. We analyzed 24 features from our ICmore and a bar plot of ICC showed that the correlation of ICmore improved after blind deconvolution in CARDIA scan-rescan (Figure 10).

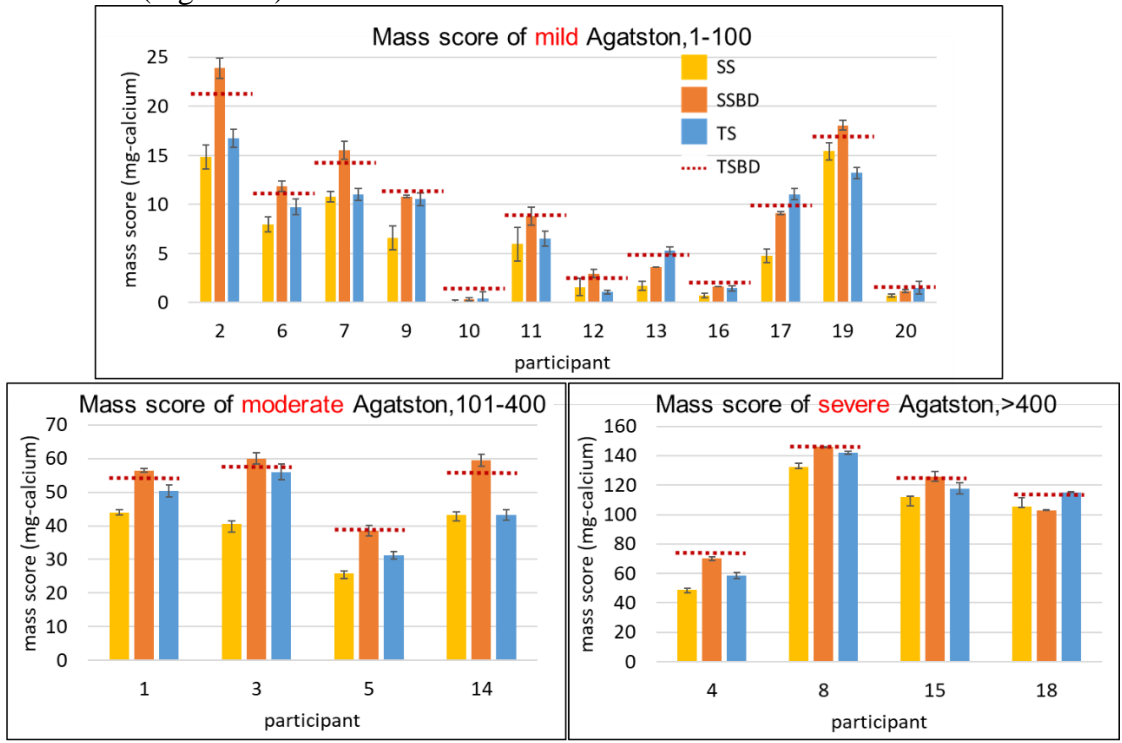

**Figure 7. Accuracy/reproducibility of CARDIA mass scores following blind deconvolution.** Image conditions (SS, SSBD, TS) are compared to TSBD(dashed). Error bars are from scan with scan-rescan. According to degree of coronary artery calcification, we divide our 20 participants into 3 groups (Agatston 1-100 (a), 101-400 (b) and above 400 (c))[37]. Similar to cadaver heart results, SSBD improved results from SS as average percent signed difference reduced from -19% to -1%. Also the averaged coefficient of variation reduced from 0.11 to 0.06.

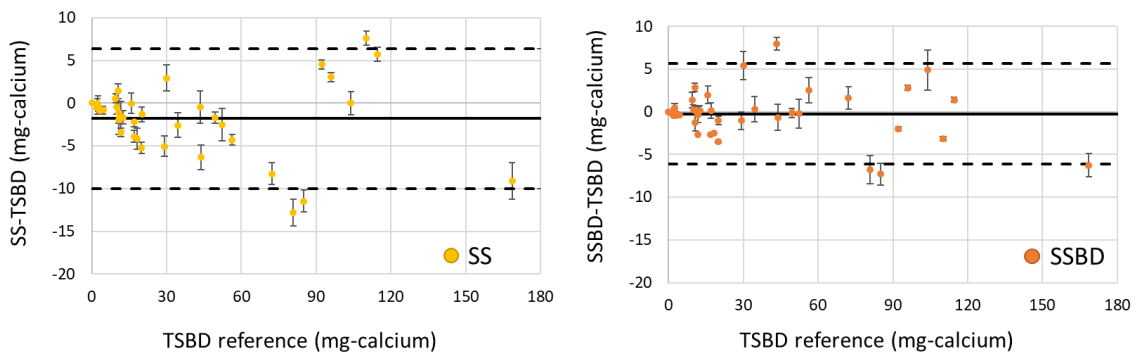

**Figure 8. Modified Bland-Altman plot of the calcium score in CARDIA participants.** We analyzed 33 calcifications in TS (1.25 mm) and SS (2.5 mm) reconstructions and compared SS and SSBD to TSBD. Similar to cadaver heart, SSBD reduces limits of agreement. Bias reduces from -1.82 to -0.27 mg-calcium, where the averaged standard deviation is comparable, from 4.2 to 6.5 mg-calcium.

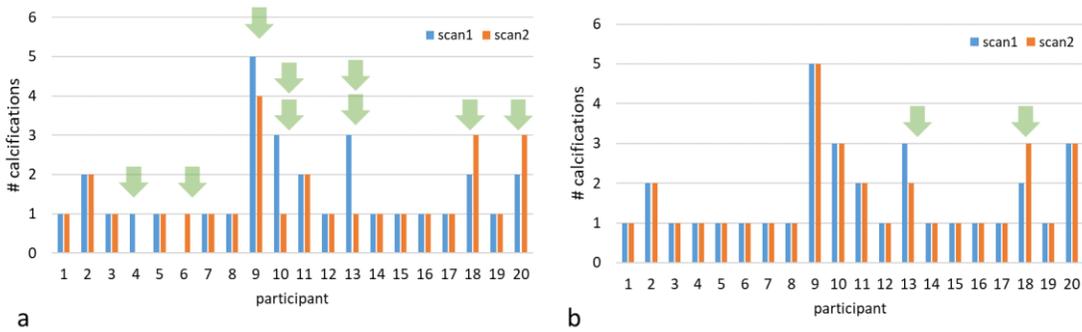

**Figure 9. Deconvolution improves reproducibility of calcification number detected in CARDIA participants.** This figure listed out number of calcification we detected in CARDIA dataset for these 20 participants. Each participant had repeated scans, the number of calcification detected were plotted as paired bars. (a) was SS analysis and (b) was SSBD. Each arrow indicated disagreement by one calcification. Deconvolution reduced differences from 9 to 2.

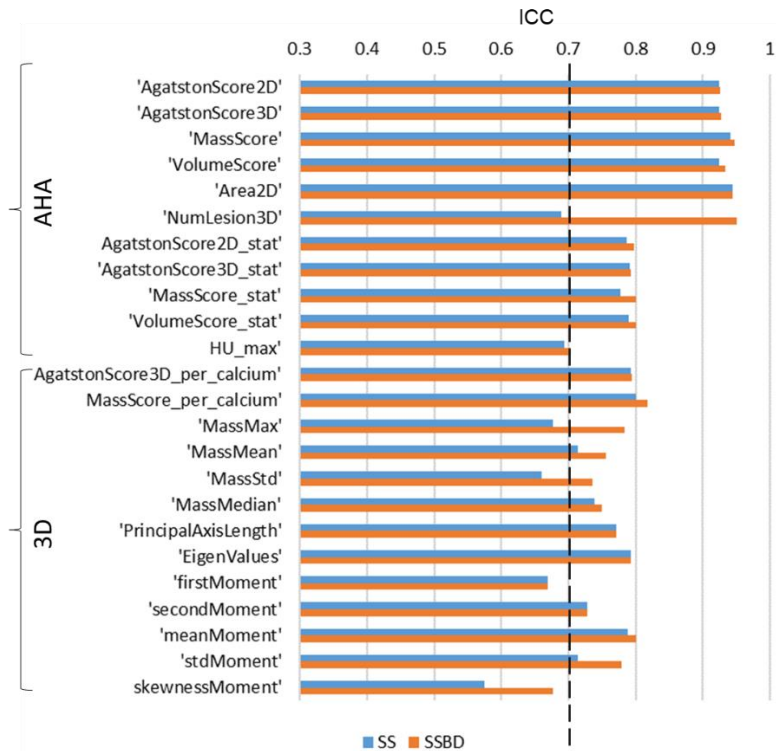

**Figure 10. Reproducibility of ICmore features in CARDIA repeated scan with and without blind deconvolution.** For 20 participants, we analyzed 24 features that were divided into Accumulated Heart and Artery (AHA) and 3D calcification. Bar plot of intraclass correlation coefficient (ICC) showed the correlation of calcification feature in CARDIA scan-rescan. "stat" is the combination of max, min, median, mean, standard deviation and "per_calcium" is individual calcification features. The vertical dashed line showed threshold of 0.7, any feature across this line showed adequate reproducibility. Blind deconvolution improved ICC in 21 out of 24 features. With blind deconvolution, 22 features were above threshold.

# 4  Discussion

In our study, we reported that deconvolution improves the accuracy and reproducibility of multiple features extracted from individual calcifications in CT calcium score exam, especially for small calcifications, which been reported to have higher risk of acute disease[13]. Further, blind deconvolution performed well comparing to conventional Lucy-Richardson deconvolution. Therefore, utilizing this technique may be beneficial for predicting adverse cardiovascular events as it provides a robust and accurate evaluation of clinically relevant and preclinical calcifications.

Feature reproducibility is a requirement for rigorous machine learning results[23–27]. To address this issue, several studies have investigated the stability of feature selection algorithms, measuring the robustness of the selected features in the data[38,39]. In the case of our results, we investigated features of the individual coronary artery calcification and traditional CT calcium score. Mass score performed the best in traditional measurements, individual calcification features like number of lesions detected significantly improved reproducibility after blind deconvolution. Blind deconvolution improved accuracy and reproducibility in our study across the features. We conclude that for archived images where PSFs are unavailable, 3D blind deconvolution is a useful preprocessing step for improved radiomics assessments of CT calcium score images.

We elegantly created ground truth data in CARDIA dataset. We selected the most approximate evaluation to ground truth in the QRM-Cardio phantom, the thin slice thickness after 3D deconvolution correction. Also we conclude that blind deconvolution showed similar performance to the 3D deconvolution based on measured PSF, proving the assumption that blind deconvolution on thin slice thickness is a reliable standard when analyzing CARDIA participants.

There are similar reports about image resolution restoration using deconvolution. Carmi et al found deconvolution significantly improved the image resolution of fine bone structures in CT images[40]. Slavine et al found deconvolution improved noisy CT image quality to potentially diagnostically acceptable levels[41]. Hehn et al investigated the feasibility of blind deconvolution on CT images and used it in conjunction with additional processing to improve results[42]. More recently, deep-learning-based methods are also being used to improve image sharpness[43]. All of these reports are aimed at improving image quality examination. In our case, we assessed quantitative improvement on individual calcification metrics towards an improved CT calcium scoring system.

## Acknowledgements

This project is partially funded from a research grant with Philips Healthcare (Exhibit B-11Y1-Coronary Calcifications). This research is a collaboration between Case Western Reserve University and University Hospitals of Cleveland. Special thanks to Mani Vembar for discussion. The Coronary Artery Risk Development in Young Adults Study (CARDIA) is supported by contracts HHSN268201800003I, HHSN268201800004I,

HHSN268201800005I, HHSN268201800006I, and HHSN-268201800007I from the National Heart, Lung, and Blood Institute (NHLBI). Analysis of CT data was in part supported by R01-HL098445, R44-HL156811, and R01-HL143484. Hao Wu was supported by the *Interdisciplinary Biomedical Imaging Training Program*, NIH T32EB007509 administered by the Department of Biomedical Engineering, Case Western Reserve University. This report is solely the responsibility of the authors and does not necessarily represent the official views of the NIH.